\documentstyle[NATO,namedreferences]{crckapb} 
\begin{opening}
\title{MEASURING THE ANISOTROPY IN THE\protect\\
       CMB: CURRENT STATUS \& FUTURE PROSPECTS }
\author{L.A. PAGE}
\institute{Dept of Physics, Princeton University\\
           Princeton, New Jersey\\
           http://physics.princeton.edu/{\char'176}page}
\end{opening}

\runningtitle{MEASURING THE CMB ANISOTROPY}

\begin{document}
\input epsf
\begin{abstract}

The CMB is perhaps the cleanest cosmological observable. Given a
cosmology model, the angular spectrum of the CMB can be computed to 
percent accuracy. On the observational side, as far as we know, 
there is little that stands in
the way between an accurate measurement and a rigorous confrontation
with theory.  In this article,
we review the state of the data and indicate future directions.

The data clearly show a rise in the angular spectrum to a peak of
roughly $\delta T_l=(l(l+1)C_l/2\pi)^{1/2}\approx 85~\mu$K at $l\approx 200$ 
and a fall at higher $l$. In particular,
$\delta T_l$ at $l=400$ is significantly less than at $l=200$. This is shown by 
a combined analysis of data sets and by the TOCO data alone.

In the simplest open models with $\Omega_m = 0.35$, one expects a
peak in the angular spectrum  near $l=400$. For spatially flat models,
a peak near $l=200$ is indicated and thus this model is
preferred by the data.
The combination of this, along with the growing body of evidence that
$\Omega_m \approx 0.3$, suggests a cosmological constant is required. 
Further evidence for a cosmological constant is provided by the height of the peak.
This conclusion is independent of the supernovae data.

\end{abstract}

\section{Introduction}

These notes are from two talks given at the Newton Institute in July
1999. The goal was to assess the status of CMB anisotropy measurements
and give some indication of what the future holds.
Given the extraordinarily rapid development of this field, this article 
is sure to be outdated soon after it appears.
The program included a section on interferometers and the data therefrom by Anthony
Lasenby, on the physics of the CMB by George Efstathiou, and on data 
analysis by Dick Bond so I shall not discuss those matters here.
Anthony Lasenby also covered work on ESA's Planck satellite. 

My talks are biased toward the experiments I know best.
Some of the experiments which I will discuss, in particular the MAT
experiments (a.k.a. TOCO97\cite{torb99} and TOCO98\cite{miller99}), 
were done by a collaboration
between Mark Devlin's group at the University of Pennsylvania and the Princeton group.

It would be stunning if the currently popular model survived
to be our favorite model of the universe in a few years.
Recent panoramic assessments of cosmological data \cite{bahcall99}
\cite{turner99} suggest the universe is made of $\Omega_b\approx 0.05$, 
$\Omega_{cdm}\approx 0.3$, \&
$\Omega_{\Lambda}\approx 0.65$\footnote{The $b$ subscript stands for
baryons, $cdm$ for cold dark matter, and $\Lambda$ for a cosmological
constant type term.}. In other words, only 5\% of the universe is
made of something with which we are familiar. 

There are three classes of observations that lead to the current picture.
The supernovae data
indicate that the universe is accelerating and thus need 
something like a cosmological constant to explain them. Secondly, the mass density, 
$\Omega_m = \Omega_b + \Omega_{cdm}$, as inferred from galactic velocities,
cluster abundances, cluster x-ray luminosities, the S-Z effect in
clusters, the cluster mass to light ratio, etc. is 
$\Omega_{m}\approx 0.35$. Thirdly, the CMB data suggest that the 
universe, within the context of adiabatic cold dark matter models,
is spatially flat \cite{bahcall99} \cite{bjk98}.
The CMB data are improving rapidly. New data (TOCO97, TOCO98, CAT\cite{bak99}), since
\cite{bahcall99}, \cite{bjk98}, strongly disfavor the nominal open spatial geometry
models, and the case is getting tighter by the month \cite{GHPJS99}.
We should point out that the position of the first peak does not {\it
prove} the universe is spatially flat; there is enough wiggle
room with the other parameters even within the limited context of 
adiabatic CDM models \cite{hu96}, but a spatially flat model is the
simplest explanation when one assumes prior knowledge of other
parameters such as $H_0$.

%
%\section{Abstract}
%If you want to make an abstract, you can use the {\stt abstract}
%environment which is standard \LaTeX. 
%You need not enter the word `Abstract'.\par
%
\section{The temperature of the CMB}

In 1990 John Mather and colleagues \cite{Mather90}, using the Far infrared
Spectrophotometer (FIRAS) aboard the 
COBE satellite, showed that the CMB is a blackbody emitter over the
frequency range of 70 to 630 GHz. It is perhaps
the best characterized blackbody. A recent analysis \cite{mather99} gives
the temperature as T$=2.725\pm 0.002$~K (95\% confidence). The error, 2 mK, is
entirely systematic and so it is difficult to assign a precise
confidence limit. 
The statistical error is of order 7 $\mu$K. This measurement was quickly
followed by the UBC rocket experiment \cite{gush90} which found 
T$ = 2.736\pm0.017~$K ($1\sigma$).

At frequencies greater than 90 GHz, the FIRAS measurement will not be bettered
without another satellite. At lower frequencies, the measurements are
less precise and there is plenty of room for improvement.
The best long wavelength measurement is by Staggs et al. \cite{staggs96} at 11 GHz.
They find $T = 2.730\pm0.014~$K. Deviations from a pure thermal spectrum are expected
to show up near a few GHz. In addition, we know the universe was reionized
at $z\approx5$, so there should be remnant free-free emission, also at a
few GHz. Unfortunately, near these frequencies, 
our Galaxy emits about 2 K making a 0.01\% 
determination of the CMB temperature difficult, to say the least.
Our picture of the spectrum of the CMB will not be complete until the
long wavelength part of the spectrum is known though only a few
groups are seriously considering these tough experiments. 

\section{The unbiased anisotropy spectrum}

Figure 1 shows all the anisotropy data that has at least made it into
preprint form (as of this writing, Oct. 99, all of it has been accepted for
publication). Many of the data points have not been confirmed or are
essentially unconfirmable; others have large calibration errors; some 
data sets comprise sets of correlated points;
still others have foreground contamination. Despite this, the
trend is clear. From the Sachs-Wolfe plateau discovered by COME/DMR
\cite{smoot92} there is a rise to an amplitude of $\delta T_l\approx 85~\mu$K
at $l\approx200$ and a fall after that.

\begin{figure}
\centerline{\vbox{\epsfxsize=8.5cm\epsfbox{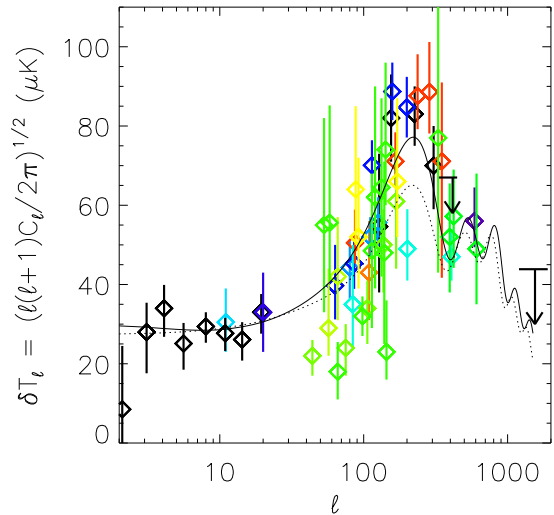}}}
\caption{Unbiased sampling of data. The solid line on top is
a model from Wang {\it et al.} (1999) with $\Omega_{b}=0.05$, 
$\Omega_{cdm}=0.3$, $\Omega_{\Lambda}=0.65$,
and $h=0.65$. The dotted curve is the ``standard cold dark matter''
model, which is inconsistent with many non-CMB observations, with
$\Omega_{b}=0.05$, $\Omega_{cdm}=0.95$, $\Omega_{\Lambda}=0.0$, and
$h=0.5$. For the PYTHON 5 data we use the data from K. Coble's thesis
(Coble 1999) rather than those from the paper.}
\end{figure}

It is worth reviewing what sort of systematic checks we have between
different experiments. At both large and small angular scales, the spectrum of the
anisotropy is seen to be thermal. Also, at the largest angular scales, there is a clear
correlation between DMR at 53 GHz and the FIRS data at 180 GHz \cite{ganga93}.
In this analysis, the dust contribution to FIRS was subtracted though
inclusion of it did not significantly alter the results: the dust is not
correlated with the CMB. At smaller angular scales, SK at 35 GHz\cite{net97} 
saw the same signal as did the MSAM experiment at 200 GHz.
In an analysis tour de force, Fixsen et al. \cite{fixsen97} show that the COBE/FIRAS 
instrument--remember FIRAS is an absolute measurement--sees the same 
anisotropy as the COBE/DMR instrument which is a
differential microwave radiometer. A plot of the cross correlation
is consistent with a thermal spectrum from 90 to 300 GHz.

Outside of the above measurements, teams have not confirmed each others'
findings. The reason is that it is difficult to match 
scan strategies from different instruments. This is one of the reasons
that maps are desired. There are preliminary indications that
the SK and QMAP maps agree \cite{doc99b} as do the maps from two seasons of 
PYTHON \cite{coblethesis}.

\begin{figure}
\centerline{\vbox{\epsfxsize=8.5cm\epsfbox{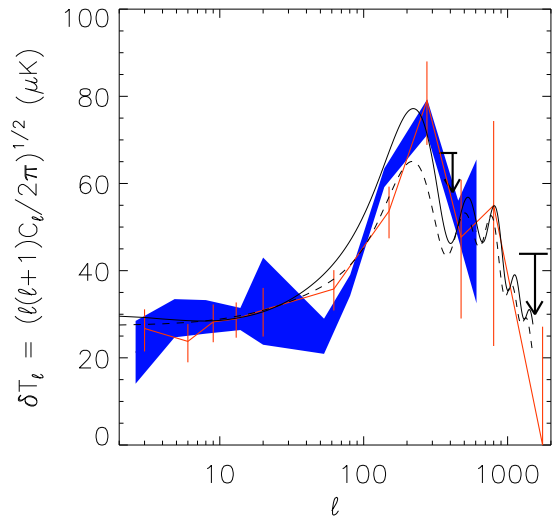}}}
\caption{All the data from Figure 1 binned into ten logarithmically
spaced bins. There is no accounting for calibration error, correlations, etc.
The upper limits are at 95\% confidence. The width of the blue swath is the 
statistical weight of the data that land in the corresponding bin. The orange line 
is a more sophisticated analysis by Bond {\it et al.} (1999) that uses a subset
of the data in Figure 1.}
\end{figure}

Figure 2 show all the data from Figure 1 binned into 
ten $l$-space bins. The plot is remarkable and gives us faith in the hot
big bang model. The rise of the angular
spectrum and the location of the peak for a spatially flat universe were 
predicted well in advance of the measurements. It is also satisfying
that these data have shown that a number of alternative models
simply do not work. For instance, large classes of isocurvature models 
do not fit the data (but by no means is the isocurvature mechanism
excluded), simple open models do not fit the data 
\cite{GHPJS99}, \cite{dodknox99}, and 
a broad class of defect models do not fit the data \cite{pen97}.

\section{The observational setting \& foreground emission}

The CMB is a 2-D random field in temperature with variance of order 
$(115~\mu{\rm K})^2$.
If one could measure the anisotropy with sharp filters in 
$l$-space, one would find the {\it rms} 
variations for $l$ between $2<l<40$ to be $54~\mu$K, between $40<l<400$ 
to be $88~\mu$K,  and between
$400<l<1500$ to be $53~\mu$K. So far, the data are consistent with a
Gaussian temperature distribution. 

Characterizing the anisotropy is challenging because one wishes to
measure accurately microkelvin variations from an experiment sitting on,
or just above, a 300~K Earth. Nature, though, has been kind.
The CMB is the brightest thing in the sky between 0.6 and 600 GHz; and
fluctuations from emission from our Galaxy (for galactic latitudes
$|b|>20^{\circ}$) are smaller than the fluctuations intrinsic to the
CMB \cite{teg99}, as shown in
Figure 3. We do not yet know if we shall be so fortunate with the
polarization, but low frequency measurements suggest this may be the
case \cite{davwil99}. 

We may get a sense of the  scale of the corrections for foreground
emission from the SK data. SK observed near the 
North Celestial Pole at $b=25^{\circ}$. The contribution to the original
data set from foreground emission is 4\% at 40 GHz \cite{doc98}. It turns out
the contamination was not due to free-free emission, Haslam-like synchrotron
emission, or extra galactic sources, but rather was due to a component 
correlated with interstellar dust emission. The favorite current 
explanation is that this component is due to radiation by spinning dust grains
\cite{dl99}. This component was not expected when the experiment was
conceived. In the QMAP experiment, the contamination at Ka ($\approx 25$
GHz ) is $\approx 8$\% \cite{doc99c}; no significant contribution to the Q band 
data was measured. Coble {\it et al.} \cite{cob99}, looking in the Southern
Hemisphere at high Galactic latitudes, found effectively no contamination at 40 GHz. 

\begin{figure}
\centerline{\vbox{\epsfxsize=8.5cm\epsfbox{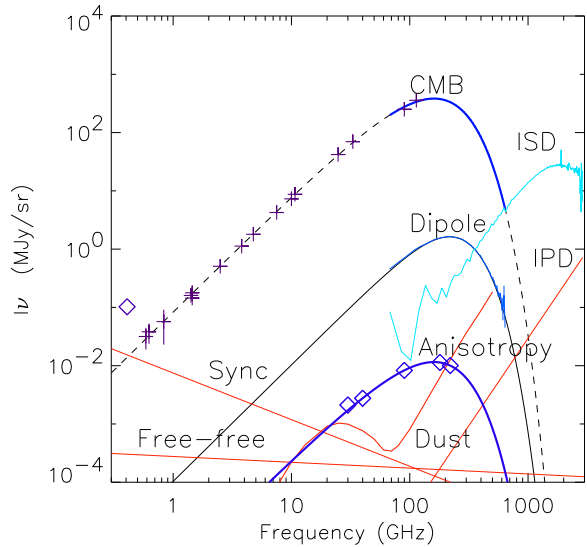}}}
\caption{Plot of the CMB and the foreground emission at approximately
the galactic latitude of the North Celestial Pole. For dust and
synchrotron spectra, the fluctuating component has been plotted (de
Oliveira-Costa {\it et al.} 1998) The flux levels of the free-free and
dust emission have also been plotted. The 
corresponding $l$ is about 20. At higher $l$, all these foregrounds have
less fluctuation power. The FIRAS data (dust, dipole, and CMB spectra) are
courtesy of Bill Reach. 
}
\end{figure}

\section{Types of measurements}

The scientific payoff from the CMB has motivated a large number of
experiments; over twenty groups are trying to measure the anisotropy.
The frequency coverage is large. The experiments and
detector technologies are summarized in \cite{barr99}, \cite{halpern99},
\& \cite{page97}. 

Here we just note that there are three general classes of measurements. 
They are (1)beam switching or beam synthesis experiments, (2)direct mapping
experiments, and (3)interferometers. By far the most data have come from
the beam switching/synthesis method but this is certain to change soon.

The detectors of choice are high electron mobility transistor amplifiers
(HEMTs \cite{posp92}, \cite{posp94}) for frequencies below 100 GHz
and bolometers for higher frequencies. The primary advantage of
HEMTs is their ease of use and speed. A typical HEMT sensitivity is 
${\rm 0.5~mKs^{1/2}}$. The advantage of bolometers is
their tremendous sensitivity, e.g. ${\rm <0.1~mKs^{1/2}}$. The CMB anisotropy 
has also been detected with SIS mixers\cite{kerr93}.

Over the past year, five new results from experiments have come out of 
which I am aware. They are the IAC\cite{dicker}, CAT, QMAP, TOCO, 
and PYTHON 5. These data
span from 30 to 150 GHz and from $l=50$ to 400. They support the picture
given in \cite{bjk98}. As an example of a beam switching experiment, I'll use
TOCO; and as an example of a mapping experiment I'll use QMAP.

\section{QMAP}

The QMAP experiment is described in a trio of papers
(\cite{dev98}, \cite{herbig98}, \& \cite{doc98}). The purpose of
QMAP, which was proposed before MAP, was to make a ``true'' map of the
sky at 30 and 40 GHz. By true map we mean a map that is simply 
described by a temperature and temperature uncertainty per pixel. Ideally,
the pixel to pixel covariance matrix is diagonal. For QMAP, this was not
the case and the full covariance matrix was required. Maps that
are reconstructed from beam switching measurements, for instance the
SK\cite{tegmark97b}, MSAM\cite{knox98}, MAX\cite{white95}, and
PYTHON 5\cite{coblethesis} maps, are not true maps and cannot be
analyzed as maps.

\begin{figure}
\centerline{\vbox{\epsfysize=5truein\epsfbox{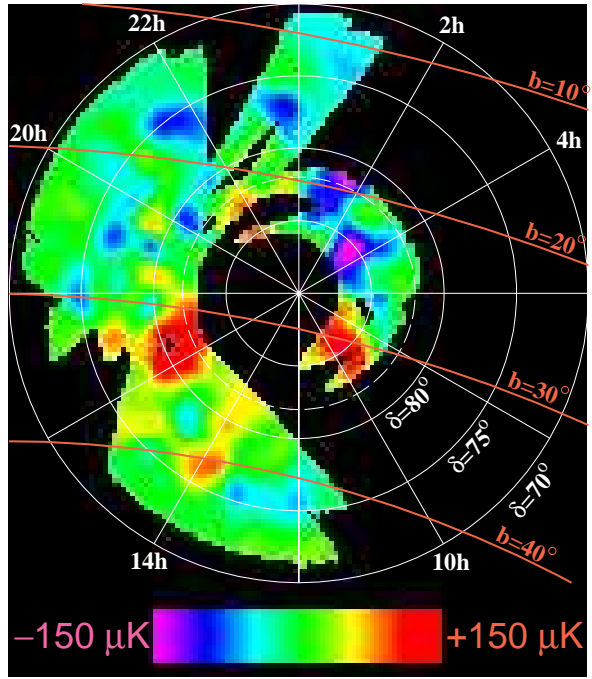}}}
\caption{Results of two QMAP flights. There are multiple spots in the map
with a signal to noise of 10 to 20. SK fills the region within
$\delta=82^{\circ}$ shown by the
dashed line. This plot shows a Weiner filtered version of the raw map.
This is what the anisotropy looks like at degree angular scales.}
\end{figure}

QMAP is a direct mapping experiment. The data stream is converted
directly into a map and the covariance matrix is computed from
the data. So far, only four experiments have 
directly mapped the CMB anisotropy (COBE/FIRAS, COBE/DMR, FIRS, and
QMAP). QMAP, so far, has the highest S/N per pixel. We can look forward
to the BOOMERanG \cite{boom} and MAXIMA \cite{maxima} data which, with
their tremendous detectors, should produce much higher sensitivity maps.

The key to making a map is to ``connect each pixel with all the ones
around it.'' \cite{tegmark97} \cite{wright96b}. In simplest terms, one wants to sit
at a pixel and know the derivatives in each direction. QMAP accomplished
this connectedness by observing above the North Celestial Pole and letting the sky
rotate through the beam. The scan lines thus intersect at a variety
of angles \cite{dev98}. Maps that are reconstructed from 
temperature differences (e.g. from beam switching experiments) do not
generally have this property because the differences
are all done at constant elevation (SK is an exception; again the
rotation around the North Celestial Pole was used.)

The QMAP power spectrum is given in Table 1. With the two flights,
and six channels per flight a number of cross checks can be made.
The data common to both flights and between channels within one flight
are consistent. In a chi-by-eye, the QMAP map looks very similar to
the reconstructed SK map in areas where they overlap \cite{doc99b}.

The QMAP data are extremely clean. There is essentially no editing
of spurious points etc. One simply takes the data, calibrates it, removes a
slowly varying offset (this is done self-consistently in the map
solution) and produces the map. It is the type of data set for which the
analysis pipeline could have been written before the experiment.

\section{MAT/TOCO}

The MAT/TOCO experiment is a collaboration between Mark Devlin's group
at Penn and the Princeton group. We took the QMAP gondola and
optics, changed the cooling from liquid helium to a mechanical refrigerator, and
mounted the telescope on a Nike Ajax radar trailer. For two seasons (Oct.-Dec. 1997 and
Jun.-Dec. 1998) we observed from Cerro Toco near the ALMA site in the 
northern Chilean Andes.\footnote{The Cerro Toco site of the Universidad
Cat\'olica de Chile was made available through the generosity of 
Professor Hern\'an Quintana, Dept. of Astronomy and Astrophysics.}

In the first season (TOCO97), all the HEMT channels 
worked but the two SIS channels
did not. The problem with the SISs was fixed for the second season (TOCO98). 
Thus we cover from $l=60$ to $l=400$ in multiple frequency bands. 
In the field we were plagued by refrigerator problems. This resulted in 
a higher SIS temperature and thus lower sensitivity than we expected 
from the laboratory measurements.

In this sort of experiment, one must deal with the variable
atmospheric temperature and variable local temperature. The data must be edited
and one must go to great measures to ensure that the editing does not
bias the answer. Of central importance is the correct assessment of the
instrument noise. 
As $\delta T_l^2$ is proportional to the measured variance minus the
instrument variance, an incorrect assessment of the noise will bias the 
result (see, for example, \cite{net97} and \cite{torb99} for
discussions). 

\begin{figure}
\centerline{\vbox{\epsfysize=8.5cm\epsfbox{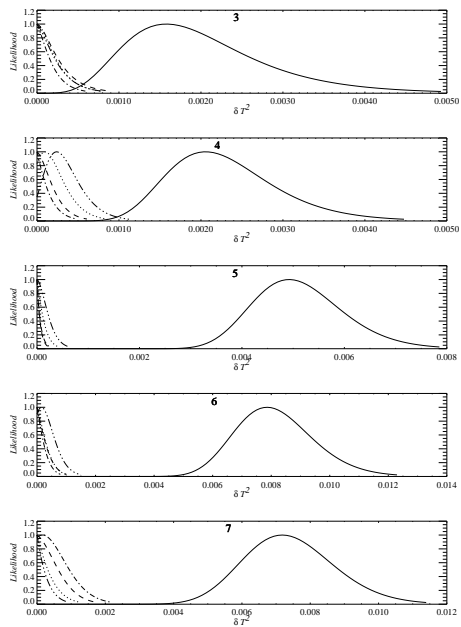}}}
\caption{The likelihood of the data and the likelihood of the null
combinations of the TOCO97 data set. The four null signals are data
taken with the chopper scanning one way minus data with the chopper
scanning the other, differences between subsequent $0.25~s$ segments,
differences between subsequent $5~s$ segments, and the first half minus
the second half of the campaign.From top to bottom, the panels
correspond to $l=63,86,114,158,$\&199 as given in Table 1.}
\end{figure}

The straight forward way to make sure that the noise is understood
is to make combinations of the data in which the sky signal is cancelled
out. The analysis of such a null signal should yield the instrument
noise. These null combinations should cover multiple time scales
and spatial scales. As an example, we show the null tests from the
TOCO97 data in Figure 5 (for TOCO98 see \cite{miller99}).
Note that in all cases, the signal is well above the instrument noise
and that for each l-space bin, the null tests, regardless of time scale,
give consistent noise levels. Additionally, the ratio of the noise between
$l$-space bins can be computed; it agrees with the data.
The results from both campaigns are in Table 1. There are roughly 100
more days of 30 and 40 GHz data to analyze.

\section{SK/QMAP/TOCO (SQT)}

I've tried to come up with an easy-to-state criteria for selecting data
sets for a compendium of solid results above
$l=50$. Qualifications for
entry onto this list would include confirmation of results, maturity of 
analyses (of order 30\% of the reported data has undergone some sort 
of reanalysis after publication resulting in significantly different
answers), some check for foreground
contamination, internally consistent data, and measurements that include 
internal consistency checks of the data quality and noise levels.
I have not been successful. As any caveated selection runs the hazard of being biased
toward results that agree with our results, I will use instead a completely subjective
criterion. Namely, experiments with which I have been involved over the
past few years. The list is given in Table 1.

The SQT data span from 30 to 150 GHz, use different
calibrators, involve different analysis packages, different radiometers,
different platforms, and different observing strategies.

Except for the QMAP point at $l=126$ which is somewhat correlated with
the other QMAP points, these data can be considered uncorrelated. 
At low $l$, all data sets are sample variance limited. The sky coverage 
varies quite a bit. For SK it is 200 deg$^2$, for QMAP 530 deg$^2$, for 
TOCO97 600 deg$^2$, and for TOCO98 500 deg$^2$.

What is the significance of the down
turn from the peak near $l=200$? The first thing to note is that the
D-band data points (TOCO98 in Figure 6) are essentially uncorrelated. There are two data
points below the maximum. The net effect is a $5\sigma$ detection of a
fall from just the TOCO98 data alone. In addition, 
including the last (low) SK point enhances the probability of a downturn.
In other words, from just these data, the down turn is indisputable.
and of course there are many other experiments as shown in Figure 1.

In Table 1 we also give the recent analysis of the MSAM data \cite{wilson99}. The
MSAM point at $l=200$ is about $3\sigma$
below the mean of the SQT points. Interestingly, the data 
comprising this point come from a section of sky examined in two MSAM flights \cite{knox98}
and on the ground in the SK experiment \cite{net97}. My interpretation 
is that this is a statistical fluke. MSAM covers only of order 10 deg$^2$ of sky. 
The experiments are well enough documented to check this against the MAP
data.

\begin{figure}
\centerline{\vbox{\epsfysize=8.5cm\epsfbox{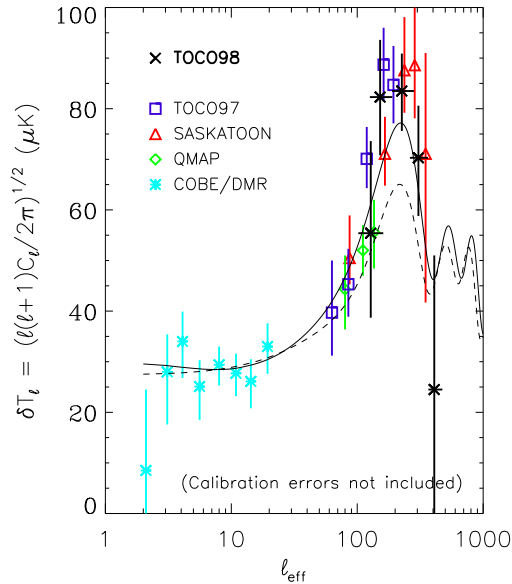}}}
\caption{The SK/QMAP/TOCO and COBE/DMR data. For the highest l point,
we show the data point using the convention of Knox. The models are the
same as in Figure 1. Note that the peak is clearly not at $l=400$.}
\end{figure}

\begin{table}[htb]
\begin{center}
\caption{Selected data. Calibrations errors ($\approx 10$\%) are not included.}
\begin{tabular}{llll}
\hline 
Name & $l_{eff}$ & $\delta T_l$ ($\mu$K) & Comments \& Reference\\
\hline
TOCO97  & $63^{+18}_{-18}$  & $39.7^{+10.3}_{-6.5}$ & Torbet et al.\\
QMAP  &$80^{+41}_{-41}$ & $44.3^{+6.7}_{-7.9}$  & Foreground subtracted \\
MSAM & $84^{+46}_{-45}$ & $35^{+15}_{-11}$ & Foreground subtracted\\
TOCO97  & $86^{+16}_{-22}$  & $45.3^{+7.0}_{-6.4}$ & Torbet et al.\\
SK & $87^{+39}_{-27}$& $50.5^{+8.4}_{-5.2}$    & Foreground subtracted\\
QMAP  &$111^{+64}_{-64}$ & $52.0^{+5.0}_{-5.0}$ & Foreground subtracted\\
TOCO97  & $114^{+20}_{-24}$ & $70.1^{+6.3}_{-5.8}$ & Torbet et al.\\
QMAP  &$126^{+54}_{-54}$ & $55.6^{+6.4}_{-7.2}$ & Foreground subtracted\\
TOCO98  & $128^{+26}_{-33}$ & $54.6^{+18.4}_{-16.6}$ & Miller et al.\\
TOCO98  & $155^{+28}_{-38}$ & $82.0^{+11.0}_{-11.0}$ & Miller et al.\\
TOCO97  & $158^{+22}_{-23}$ & $88.7^{+7.3}_{-7.2}$ & Torbet et al.\\
SK & $166^{+30}_{-43}$& $71.1^{+7.3}_{-6.3}$   & Foreground subtracted \\
TOCO97  & $199^{+38}_{-29}$ & $84.7^{+7.7}_{-7.6}$ & Torbet et al.\\
MSAM & $201^{+82}_{-70}$ & $49^{+10}_{-8}$ & Foreground subtracted\\
TOCO98  & $226^{+37}_{-56}$ & $83.0^{+7.0}_{-8.0}$ & Miller et al.\\
SK & $237^{+29}_{-41}$& $87.6^{+10.5}_{-8.4}$  & Foreground subtracted \\
SK & $286^{+24}_{-36}$& $88.6^{+12.6}_{-10.5}$ & Foreground subtracted \\
TOCO98  & $306^{+44}_{-59}$ & $70.0^{+10.0}_{-11.0}$ & Miller et al.\\
SK & $349^{+44}_{-41}$& $71.1^{+19.9}_{-29.4}$ & Foreground subtracted \\
MSAM & $407^{+46}_{-123}$ & $47^{+7}_{-6}$ & Foreground subtracted\\
TOCO98  & $409$ & $<67~(95\% conf)$ & Miller et al\\
\hline
\end{tabular}
\end{center}
\end{table}

\section{Finding the peak}

Once we have the power spectrum we can either fit the data to models
or we can look for nearly model independent parametrizations. Directly
fitting to models has been done by a number of groups
(e.g.  \cite{bahcall99}, \cite{bartlett98}, \cite{bond98b},
\cite{dodknox99}, \cite{line99}, \cite{ratra99}, \& \cite{tegmark99b})
In broad brushstrokes, the data are consistent with spatially
flat models with a cosmological constant and inconsistent with
spatially open models (though this conclusion depends on the selection
of data sets \cite{ratra99}). In the following, we give a model independent
assessment of the position of the peak.

The richness of the CMB is the very thing that makes it so difficult to
fit. In the current stage, one cannot look at the spectrum and get
simple answers. There are multiple sets of parameters that give rise to 
the same power spectrum if only a certain region of $l$-space is covered 
\cite{bond94}. This degeneracy is broken by including other data sets into
the analysis; for instance, one may assume prior knowledge of
the Hubble constant or the baryon density.

In an effort to say where the peak is, we have parametrized a generic
spectrum by taking the lambda model in Figure 1\cite{wang99}, normalizing it to 
$\delta T_l=32$ at $l=25$ and stretching it in $l$ and changing the
amplitude while maintaining the normalization. (To my knowledge, this 
was done first by Barth Netterfield to the SK data. At that time, we 
found that the SK peak preferred h=0.35, a cosmological constant, or
lots of baryons. Basically, anything to make the peak higher than the
sCDM peak.)
We then compute the likelihood as a function of $l$ 
and amplitude of the peak. The results are shown in Figure 7
for both the SQT and all the data in Figure 1.

\begin{figure}
\centerline{\vbox{\epsfysize=8.5cm\epsfbox{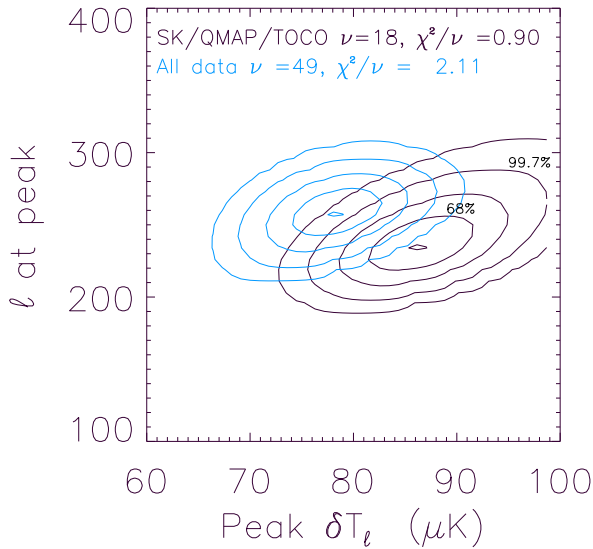}}}
\caption{The peak position and amplitude for the SK/QMAP/TOCO data 
and all the data in Figure 1. DMR was not included. Note that the 
reduced chi-squared for SQT is completely consistent with statistical
uncertainty. Note also that a peak near $l=400$ is very unlikely. 
Calibration error shifts the contours left and right. }
\end{figure}

To show what the data prefer, we have done a simple 
$\chi^2$ analysis assuming a spatially flat universe and 
limiting ourselves to variations of $\Omega_{\Lambda}$, $\Omega_{b}$ 
and $\Omega_{cdm}$ with h=0.65.
The SQT data are well described by
$\Omega_{b}=0.05$, $\Omega_{cdm}=0.2$, \& $\Omega_{\Lambda}=0.75$.
If one wants to explain the height of the peak without a cosmological
constant, then one must have something like $\Omega_{b}=0.12$,
$\Omega_{cdm}=0.88$, clearly at variance with a large body of cosmological data.

We have reached the point where there are well developed techniques for
measuring and quantifying the anisotropy from balloons and the ground 
that give reliable and consistent results. This situation will continue
to improve with the large interferometers and highly sensitive
balloon data coming on line now. The current limitations to the
experiments are 1)calibration, 2)sky coverage and 3)knowledge of the beam.
The dominant error for the combined SQT data is the calibration.
We find that the amplitude of the peak between $l=150$ and $l=250$ 
is $\delta T_l=82\pm3.3\pm5.5~\mu$K, the first error is statistical and
the second error is from the calibration. A typical intrinsic calibration source
accuracy is 5\%. Secondly, without full sky coverage, the ultimate error
bars cannot be achieved. For instance, if one maps only a quarter
the full sky, the error bars will be twice as large as potentially achievable.
Finally, there is no reason that the
beams cannot be known to high accuracy. The problem is that they must be
measured {\it in situ} and these measurements can be difficult. 
To the accuracy needed, beams can no longer simply be modeled
by two dimensional Gaussian profiles.

\section{The past}

It is amusing to look back and see how far we have
come in the past six years. Figure 8, is a plot of all the data 
that were published in or before 1993 with sensitivities at an
interesting level. I've converted the reported limits on the Gaussian
autocorrelation function to the modern band powers using \cite{bond94a}
and \cite{gundthesis}. The full power spectrum of DMR had not been published
at this time. Based on these data, one would be surprised by the models
that currently fit so well.

\begin{figure}
\centerline{\vbox{\epsfysize=8.5cm\epsfbox{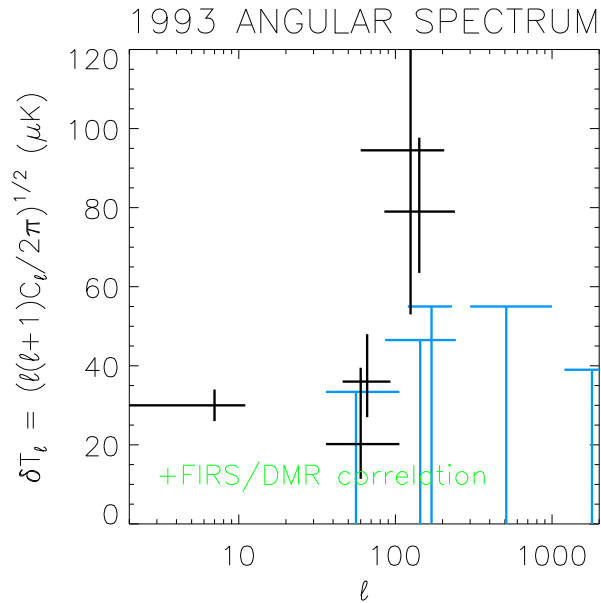}}}
\caption{Plot of the CMB angular spectrum of data published in 1993 and
before. The upper limits are all at 95\% confidence. No measurement
between $l=100$ and $l=1000$ has been confirmed. }
\end{figure}

There have been tremendous advances on the theoretical front as well. 
Model predictions have been brought to the masses through CMBFAST
\cite{selzal}; textures and similar mechanisms are generally 
believed to be inconsistent 
with the current data \cite{pen97}, and there are many classes of 
isocurvature models that are no longer viable. As the theories improve, 
some of these models may arise
again or we may find that the anisotropy is an admixture of
adiabatic, isocurvature, and texture perturbations.

\section{The future}

At $l<1000$ the future is in multielement interferometers \cite{CBI}
\cite{DASI}, 
long duration balloon flights e.g (\cite{boom}), and satellite
missions. At $l<20$, the sky can only be mapped precisely from a 
satellite. For $l>1000$, measurements can be made from the
ground. Arrays of bolometers and interferometers seem ideal. The
polarization has yet to be detected but experiments coming on line now
should be able to do the job within a year or so.

There are now four space missions on the books. There is NASA's MAP
satellite which just had its Launch-1 year review, there is ESA's Planck
satellite (with collaborative NASA support) which is scheduled for a 2007
launch, there is the SPORT mission which plans to measure the
polarization of the CMB at HEMT frequencies from the space 
station \cite{cortiglioni99}, and, in NASA's technological road map, there is a
mission, CMBPOL, to measure the polarization of the CMB in $\approx 2015$. The later
is in the talking phase \cite{pet99}. I shall focus on MAP and 
Anthony Lasenby will focus on Planck.
 
\section{MAP}

The primary goal of MAP\footnote{MAP is a collaboration between 
NASA (Chuck Bennett [PI], Gary Hinshaw, Al Kogut, \& Ed Wollack), 
Princeton (Norm Jarosik, Michele Limon, Lyman Page, David Spergel \& David
Wilkinson), Chicago (Steve Meyer), UCLA (Ned Wright), UBC (Mark
Halpern), and Brown (Greg Tucker)} is to produce a high fidelity, polarization
sensitive, full sky map of the cosmic microwave background anisotropy.
From its inception, the focus has been on how one makes a full sky map with
negligible systematic error. MAP is a MIDEX mission which means there
is minimal redundancy as well as firm cost and schedule caps. 
MAP was proposed in June 1995, selected in April 1996, and is planned for
a November 2000 launch. It was also proposed with the notion of getting 
the data to the community as fast as possible. We plan to make maps 
public nine months after we
scan the whole sky (roughly one year after getting to L2).

Sources of systematic error in mapmaking include ``1/f noise'' (any 
variations from tenths of seconds to minutes) in the detectors and
instrument, magnetic fields, and sidelobe contamination. Our systematic 
error budget allows for a total of $5~\mu$K of extraneous
signal {\it before} any modeling of the source of the contamination. 
The $5~\mu$K applies to all angular scales and time scales though
the most troublesome sources are the ones that are synchronous with the spin period.

MAP is being documented on the web as the instrument is built and
tested. Official information may be obtained at
http://map.gsfc.nasa.gov/; technical papers are in preparation.

\begin{figure}
\centerline{\vbox{\epsfysize=4truein\epsfbox{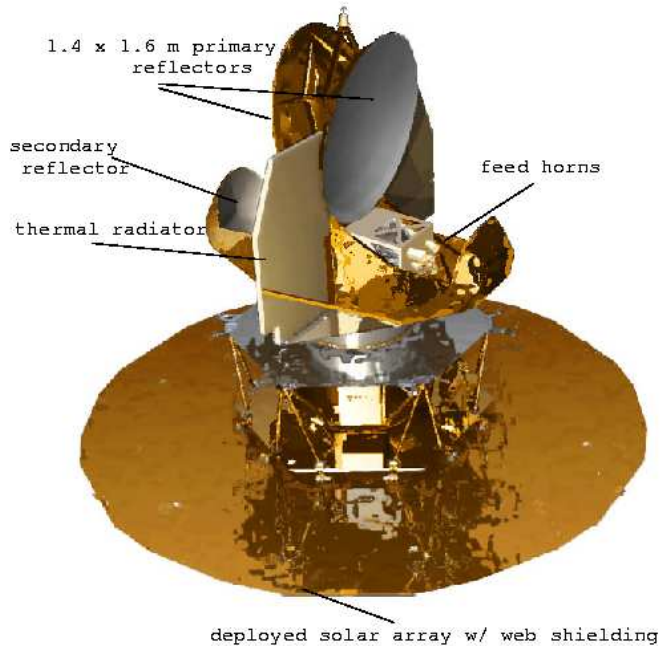}}}
\caption{Picture of MAP with the solar arrays deployed. The receivers
are located directly beneath the primary mirrors and are cooled by the
thermal radiators. The Earth, Moon, and Sun are beneath the solar arrays.}
\end{figure}

\subsection{Mission outline}

The design guidelines for MAP were:

\begin{itemize}
\item {\bf Simplicity} Other than the thruster valves and 
attitude control reaction wheels, there are no moving parts when the satellite is at L2. 
For taking science data, there is only one mode of operation. MAP
is passively cooled to $\approx 95~$K.
\item {\bf Stability} The orbit at the Earth-Sun Lagrange point, L2, is
thermally stable and has a negligible magnetic field. L2 is roughly
$1.5\times10^{6}$ km from Earth and the 
Sun, Earth, and Moon are always ``under'' the spacecraft.
(It will take approximately 3 months to get to L2.)
\item {\bf Heritage} No major new components were required except the NRAO W-band
amplifiers. These were designed for MAP.
\item {\bf Ease of Integration} The mission relies on the complete
understanding of the systematic noise levels. The magnitude of many of the
systematic effects are easily determined
when the instrument is warm. 
\end{itemize}

with spacecraft and mission parameters as follows:

\begin{itemize}
\item {\bf Mission Duration} Two years at L2. It can last longer.
\item {\bf Mass} 800 kg
\item {\bf Power} 400 W
\item {\bf Launch Vehicle} Delta 7425 (4 strap on motors).
\end{itemize}

\subsection{Receivers}

MAP uses pseudo-correlation radiometers to continuously measure the
difference in power from two input feeds on opposite sides of the 
spacecraft. The detecting elements are HEMTs designed by Marian
Pospieszalski at the National Radio Astronomy Observatory (NRAO).
The wide bandwidth and high sensitivity of the HEMTs, even while 
operating near 95~K, are what make MAP possible.
There are ten feeds on each side of the spacecraft and each feed
supports two polarizations, thus there are a total of twenty
differential chains. Each receiver chain uses four amplifiers (two
at $\approx 95$~K and two at $\approx 290$~K) for a total of eighty amplifiers.
The radiometers are configured so that they difference two polarizations
whose electric fields are parallel.

The key characteristic of the receivers is their low 1/f noise. 
The noise level at the spin rate, 8 mHz, is virtually the same
as at the 2.5 kHz switching rate. This means that the receivers will not
correlate noise from one pixel to the next. To put this receiver performance in
perspective, we note that the 1/f knee of the W-band HEMTs alone is near 1 kHz.
There is a complete model of the receivers from the feeds to the 
detector outputs, including all the support electronics. 
With the model, and measurements that correlate the model to reality, 
we determine the sensitivity of the output to temperature 
variations in each component. The temperatures of the most sensitive 
components are monitored during flight to roughly 1 mK accuracy.

Due to the wide HEMT bandwidth, the central effective frequency of the band
depends on the source one is observing. Between dust and synchrotron
sources, the shift is greater than 1 GHz in W-band. This will have to be
taken into account when observing different sources. Below is a table 
of the representative central frequencies and noise bandwidths of the 
receivers. The values in flight will be different and are channel dependent.

\begin{table}[htb]
\begin{center}
\caption{MAP band centers and noise bandwidth}
\begin{tabular}{cccc}
\hline 
Band & $f_c$ (GHz)& Bandwidth (GHz) & \# Channels\\
\hline
K & 23 & 5.3 & 4  \\
${\rm K_a}$ & 33 & 7 & 4 \\
Q & 41 & 8.4 & 8 \\
V & 61& 12 & 8\\
W & 95& 17 & 16\\
\hline
\end{tabular}
\end{center}
\end{table}

\subsection{Optics}

The optics comprises two back-to-back telescopes. The secondary 
of each telescope is illuminated by a cooled corrugated feed
which is the input to the receiver chain.
The reflector surfaces are shaped to optimize the beam profiles but to
a good approximation, the telescopes are Gregorian.
The Gregorian design was chosen because it is more compact than
the Cassegrain and because it could accommodate the back-to-back
feed geometry required for the radiometers. We considered a wide variety
of designs, including single and triple reflector systems, but the
Gregorian suited our needs best.

The feeds and receivers occupy a large space. As a consequence the beam profiles are
not symmetric (see Table 3). The scan strategy, to a first approximation,
symmetrizes the beam profile. Thus even if the beams were 2-D Gaussians
to start with, the symmetrized profile would not be Gaussian. Likewise,
the window functions are not simply Gaussian. Nonetheless, the beam profiles
and windows can be parametrized by Gaussians for most work. The data in
the table are representative, the final beams will be measured in flight
with sub-percent accuracy.

\begin{table}[htb]
\begin{center}
\caption{Approximate E \& H plane beam $\theta_{\rm FWHM}$}
\begin{tabular}{ccc}
\hline 
Band & $\theta_E$ (deg)& $\theta_H$ (deg) \\
\hline
K & 0.95 & 0.75    \\
${\rm K_a}$ & 0.7 & 0.6 \\
Q & 0.45 & 0.5  \\
V & 0.3& 0.35 \\
W & 0.21& 0.21 \\

\hline
\end{tabular}
\end{center}
\end{table}

Outside of making the optics fit into the MIDEX fairing, one wants to
ensure that one measures power from only the main beam. The Sun, Earth, 
and Moon, which are always at least $100^{\circ}$ away from the main
beam, are blocked by the solar arrays. The contribution from these
sources is computed to be much less than $~1\mu$K. The more difficult source to
block is the Galaxy which illuminates the feeds from just over the top
of the secondary. To block it, we substantially 
oversized the secondary, so that at its edge the illumination from the 
feed is less that $10^{-5}$ the illumination at the center ($<-50~$dB edge taper).  

The optics are modeled using a program from YRS Associates\cite{yrs} that solves
for the currents on the reflectors as waves propagate through the
system. The measured beam profiles (all have been measured at ten
frequencies across the band) are in excellent 
agreement with the predictions. Chris Barnes modified the code to run on
a supercomputer so that we can also predict the sidelobes. The sidelobe
measurements agree well with the predictions. 

Using the models, we can estimate the contribution to MAP from the
Galaxy. Our model includes a spinning dust component so that
the frequency spectrum resembles something like that shown in \cite{dl99}. We then
fly MAP over the Galaxy and record two signals. One signal is the rms
difference between the two telescopes with the response integrated over the
whole sky; the second signal is the same except with the contribution
from the main beams subtracted. In other words, the second method tells
how much Galactic signal comes through the sidelobes or from angles
greater than $\approx 4^{\circ}$ from the main beam. We do this for
$|b|>15$. The model is only approximate and does not yet include a
contribution from extragalactic
sources. Of course, the model will be updated after we measure what 
the Galaxy is really like.

\begin{table}[htb]
\begin{center}
\caption{Approximate Galactic contributions for $|b|>15^{\circ}$}
\begin{tabular}{ccc}
\hline 
Band & Galactic contribution ($\mu$K)& Sidelobe contribution ($\mu$K) \\
\hline
K & 120 & 16  \\
${\rm K_a}$ & 60 & 2 \\
Q & 40 & 4  \\
V & 20 & 0.2 \\
W & $\cdots$ & $\cdots$\\
\hline
\end{tabular}
\end{center}
\end{table}

To put these numbers in perspective, the {\it rms} magnitude of the CMB is
about $120~\mu$K. In K band, the {\it rms} Galaxy signal is roughly the same. 
To first order, these add in quadrature to produce a signal with an
{\it rms} of $170~\mu$K. In V-band, the galactic contribution is far
less; it will change the power spectrum by $\approx 2$\% if uncorrected.
These rough numbers show that the power spectrum estimates are quite robust to
the Galactic contamination. 

As we have emphasized, the goal of MAP is to make maps; the
power spectrum is just one way to quantify them. In producing a map of
the CMB, we will clearly have to model and subtract the Galaxy. The rightmost column
in Table 4 gives an indication of the contribution to the map if the sidelobe
contribution is not accounted for.

\subsection{Scan strategy}

The scan strategy is at the core of MAP and can only be realized at a place
like L2. To make maps that are equally sensitive to large and small
scale structure, large angular separations must be measured with small
beams. To guard against variations 
in the instrument, as many angular
scales as possible should be covered in as short a time as possible. 
 MAP spins and precesses like a
top. There are four time scales. The beams are differenced at 2.5 kHz;
the satellite spins at 0.45 rpm; the spin axis precesses around a 
$22.5^{\circ}$ half angle cone every hour; and the sky is fully covered 
in six months. In one hour, roughly 30\% of the sky is mapped.

As of this writing, the core of a pipeline exists to go from the time ordered
data to maps to a power spectrum.

\subsection{Science}

Of primary interest, initially, will be the angular spectrum. It will be
calibrated to percent accuracy and there will be numerous independent internal
consistency checks. Of central importance will be the accompanying 
systematic error budget. The power spectrum will be sample variance
limited up to $l\approx 700$. In other words, if the systematic errors
and foreground contributions are under control, it will not be possible 
to determine the angular spectrum any better. MAP will be sensitive up 
to $l\approx 1000$.

The angular spectrum is not the best metric for assessing MAP. The
primary goal
is a map with negligible correlations between pixels. With such a map,
analyses are simplified and the map is indeed a true picture of the sky.
MAP should be able to measure the temperature-polarization cross 
correlation \cite{crit95} and will be sensitive to the polarization signal itself.
Correlations with X-ray maps will shed light on the extended Sunyaev Zel'dovich effect 
(not to mention the dozen or so discrete sources\cite{ref98}). Correlations with
the Sloan Digital Sky Survey will inform us about large
scale structure.  MAP will be calibrated on the CMB dipole
and thus will be able to calibrate radio sources and planets to a universal system. It
will also help elucidate the emission properties of the 
intergalactic medium.

\section{Conclusions}

This is a truly amazing era for cosmology. Our theoretical
knowledge has advanced to the point at which definite and testable
predictions of cosmological models can be made. For the CMB
anisotropy, results from the angular power spectrum, frequency
spectrum, statistical distribution, and polarization must all be 
consistent. In addition, these results must be consistent  with
the distribution, velocity flows, and masses of galaxies and
clusters of galaxies as well as the age of the universe. There is 
a fantastic interconnecting web of constraints. 

Using the CMB to probe cosmology is still in its early phase. After all,
the anisotropy was discovered less than a decade ago. As cosmological models
and measurements improve, the CMB and other measures will become a tool 
for probing high energy physics. For example, should the cosmological 
constant survive, we will have a handle on new physics in the early
universe that we could never have obtained from accelerators.

\section{Acknowledgements}

I thank Chuck Bennett, Mark Devlin, Amber Miller, Suzanne Staggs, and Ned Wright for 
comments that improved this article. In the course of writing, a paper describing the  
VIPER experiment\cite{pet99b} appeared. These data are not included in
the analysis nor do they change any conclusions.

\end{document}